\documentclass{optica-article}

\journal{opticajournal}

\articletype{Research Article}

\usepackage{lineno}

\usepackage{graphicx}%
\usepackage{multirow}%
\usepackage{amsmath,amssymb,amsfonts}%

\usepackage{amsthm}%
\usepackage{mathrsfs}%
\usepackage[title]{appendix}%
\usepackage{xcolor}%
\usepackage{textcomp}%
\usepackage{comment}
\usepackage{manyfoot}%
\usepackage[symbol]{footmisc}
\usepackage{booktabs}%
\usepackage{algorithm}%
\usepackage{pdfpages}
\usepackage{algorithmicx}%
\usepackage{algpseudocode}%
\usepackage{listings}%

\begin{document}

\title{Generating quantum entanglement from sunlight}

\author{Cheng Li,\authormark{1,$\dagger$} Jasvinder Brar,\authormark{2,3,$\dagger$} Michael Küblböck,\authormark{2,3} Jeremy Upham,\authormark{1} Hanieh Fattahi,\authormark{2,3,*} and Robert W. Boyd\authormark{1,4,*}}

\address{\authormark{1}Department of Physics, University of Ottawa, Ottawa, Ontario K1N 6N5, Canada\\
\authormark{2}Max Planck Institute for the Science of Light, Erlangen, Germany\\
\authormark{3}Friedrich-Alexander-Universität Erlangen-Nürnberg, Erlangen, Germany\\
\authormark{4}Institute of Optics, University of Rochester, Rochester, New York 14627, USA\\
\authormark{$\dagger$} These authors contributed equally to this work.}

\email{\authormark{*}hanieh.fattahi@mpl.mpg.de, \authormark{*}rboyd@uottawa.ca}

\begin{abstract}
Energy consumption is on track to become a serious bottleneck in integrating quantum technologies with existing information and communication infrastructure. In photonic quantum technologies, a considerable portion of the energy overhead stems from the use of lasers, whose high coherence has long been considered indispensable for preparing quantum states of light. Here, we demonstrate that despite their low optical coherence, natural sources of light, such as sunlight, can produce quantum-entangled photon states by means of nonlinear optical effects. From spontaneous parametric down-conversion pumped by sunlight, we detect polarization-entangled photon pairs with a concurrence of $C =0.905\pm0.053$, a purity of $P =0.919\pm0.045$, and a fidelity to a Bell state of $F = 0.939\pm0.027$. The resulting two-photon state violates Bell's inequality with $S =  2.5408 \pm 0.2171$, which exceeds the threshold value of 2 to confirm quantum behavior. More importantly, we observe a photon pair generation rate of $\sim 1600~\mathrm{s}^{-1}~(\mathrm{mW~of~pump~power})^{-1}$, which is comparable to that of the laser-pumped SPDC when normalized for the spectral bandwidth of the pump. Our work paves the way for developing sustainable and accessible photonic quantum technologies, especially for implementation in resource-constrained areas such as the Arctic region and satellites in space for interplanetary missions.
\end{abstract}

\section{Introduction}\label{sec1}

Quantum technologies have promised to revolutionize modern information processing, offering computational advantages for classically hard problems~\cite{feynman1982qc, shor1994, kahanamoku2022natphy}, information-theoretic communication security~\cite{Bennett1984, Ekert1991PRL, portmann2022revmodphy}, and sensing precision beyond classical shot-noise limits~\cite{Giovannetti2004science, degen2017RevModPhy}. Over the past few decades, breakthroughs in device performance have been widely celebrated in both scientific journals and the popular press, while a critical figure of merit for these advances---the energy cost---has raised concerns for their sustainability~\cite{auffeves2022quantum}. In the classical domain, the global information and communication technology sector already accounts for $\sim$ 1.8–3.9\% of worldwide greenhouse gas emissions and energy use~\cite{itu2023ict, freitag2021patterns}. With the ever-growing global data traffic, any large-scale deployment of quantum-enabled information and communication technology will inevitably add to this burden. 

On existing quantum platforms, a substantial share of the power consumption stems from the energy overhead required to prepare and control the physical quantum systems. Preserving quantum coherence in superconducting circuits requires millikelvin-level cooling, which is provided by cryogenic equipment that routinely consumes 5-10 kW of electrical power per unit~\cite{oxinst_dilution, parker2023energy}. Ion-trap processors require energy-intensive resources such as ultra-high-vacuum infrastructure and high-power radiofrequency drive fields~\cite{kielpinski2002nature}. In photonic quantum systems, generating quantum states of light often involves deploying lasers, which have suboptimal electrical-to-optical conversion efficiency due to their inherent requirement of operating above a driving current threshold. Commercial lasers typically draw watts of electrical power only to deliver optical power on the milliwatt scale, with a significant portion of the consumed energy going towards spectrum stabilization and temperature control, or dissipating as heat. Moreover, their durability and long-term reliability are limited when operating in harsh conditions such as intense radiation, high vacuum, and extreme temperature fluctuations. The energy requirements for these platforms could significantly constrain the scalability and accessibility of quantum technologies. Fortunately, operations of photonic quantum systems may not require enduring the energy burden from lasers, unlike superconducting and ion-trap systems, whose underlying physical principles necessitate the energy overheads. 

A key ingredient in photonic quantum technologies is the entangled photon state~\cite{Jennewin2000PRL, Ursin2007NatPhysics, Black2019PRL, Yin2020Nature, defienne2021natphys, Cameron2024Science}. Lasers often serve as the pump source for generating entangled photons from spontaneous parametric down-conversion (SPDC), which is a nonlinear optical process widely employed in many photonic quantum applications. In SPDC, photons from a pump beam interact with a nonlinear medium and are down-converted into photon pairs~\cite{Burnham1970PRL, Hong1985PRA, Boyd2020NLO}. Conventionally, lasers are employed as pump sources due to their typically high coherence across all degrees of freedom, including spatial, spectral-temporal, and polarization. The reason is that the coherence of the pump in a given degree of freedom puts an upper bound on the maximally attainable entanglement in the same degree of freedom \cite{Jha2010PRA, Giese2018PhysicaScripta, Monken1998PRA, Hugo2019PRA, Zhang2019OptExpress, Burlakov2001PRA, Jha2008PRA, Kulkarni2017JOSAB, Kulkarni2016PRA, Meher2020JOSAB}. However, this constraint does not apply when the target entanglement is in a different degree of freedom from that of the coherence of the pump~\cite{Hutter2020PRL}. For instance, studies from the authors, as well as others, have demonstrated that a polarized but spatiotemporally incoherent light source, such as a light-emitting diode (LED), can produce polarization-entangled photon pairs via SPDC \cite{Li2023PRA, Zhang2023PRApplied, li2025pra}. These earlier achievements suggest that natural sources of light, such as sunlight, can replace lasers as the pump source for nonlinear optical processes and generate entangled photons, thereby enabling energy-efficient quantum technologies.

Solar energy harvesting through photovoltaics offers a prominent option for green, renewable power supply. However, the efficiency of state-of-the-art solar cells remains below $50\%$~\cite{nrel_cell_efficiency}, and their performance and durability are similarly constrained in severe environmental conditions. In parallel with the materials science effort in solar cell research, sunlight concentration technologies have been developed over the years to facilitate optimal utilization of solar power~\cite{tian2018cpc}. Studies have proposed using solar concentration technology to directly drive optical amplification and build a solar laser~\cite{holloway1981comparative, kiss1963sun, kueblboeck2024solar}. On the other hand, large-scale solar-powered space infrastructure, such as space-based solar power platforms~\cite{Glaser1973SolarPatent} and orbital data centers~\cite{Aguera2025SpaceAI}, is rapidly advancing, driving interest in photonic technologies capable of operating directly on abundant, incoherent solar radiation without dependence on power-intensive laser or electrical subsystems. Within this landscape, solar-driven photonic quantum technologies represent a promising route toward energy-autonomous quantum photonic functionality, naturally aligned with the requirements of future space power and information architectures. 
Here, we demonstrate, for the first time, that sunlight, a ubiquitous and environmentally friendly light source, can produce entangled photon pairs via SPDC despite its lack of optical coherence. Using a combination of a Fresnel lens and a conic concentrator, we couple the sunlight into a multimode fiber (MMF) that guides the sunlight into a free-space optical entanglement source setup. The spectrally filtered sunlight then pumps a nonlinear crystal placed inside a polarization Sagnac interferometer (PSI) to induce SPDC \cite{Kim2006PRA}. We measure photon coincidences with a time correlation width of $\sim1$~ns. By conducting quantum state tomography\cite{James2001PRA}, we measure a two-photon state with a concurrence of $C =0.905\pm0.053$ \cite{Wootters1998PRL}, a purity of $P =0.919\pm0.045$, and a fidelity to the target Bell state of $F = 0.939 \pm 0.027$. The polarization correlations between photon pairs result in a Bell parameter for a Clauser-Horne-Shimony-Holt (CHSH)-type measurement of $S =  2.5408 \pm 0.2171$~\cite{Bell1964, Clauser1969PRL}, which surpasses the local realistic threshold of $S=2$ by 2.49 standard deviations, indicating that the correlation cannot be fully explained by a local hidden variable theory. These results are not simply a proof of principle that polarization entanglement is generated from SPDC pumped by sunlight; they are a quantitative demonstration that these photon pairs have entanglement close to that of many recent results on polarization entanglement generated from SPDC pumped by coherence laser sources~\cite{Lohrmann2020APL, Lee2021QST, Brambila2023OE, Park2025apr}. We attribute the non-maximal entanglement and purity to practical imperfections in the experimental conditions, such as the wavefront distortions introduced by optical components, not to any inherent issue with the coherence of our solar pump. More importantly, we observe that the generation rate of entangled photons from sunlight-pumped SPDC, which is $\sim 1600~\mathrm{s}^{-1}~(\mathrm{mW~of~pump~power})^{-1}$, is comparable with that of laser-pumped SPDC when normalized against the effective phase-matching bandwidth. Our proof-of-principle demonstration not only indicates that there is no physical impediment to generating entangled photon pairs from incoherent sunlight, but is also highly encouraging that technical optimization of sunlight concentration at the phase-matching wavelengths of SPDC could make it competitive with current laser-pumped sources of polarization entangled photon sources. Our results serve as an important step towards solar-driven photonic quantum technologies and have important implications for developing energy-efficient and environmentally friendly quantum devices. 

\begin{figure}[ht]%
\centering
\includegraphics[width=0.9\textwidth]{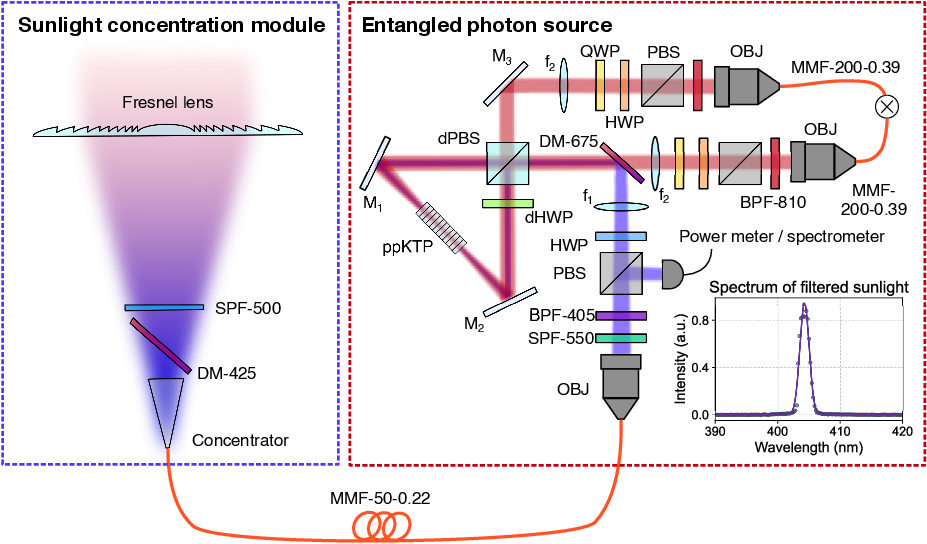}
\caption{\label{fig:1} \textbf{Schematics of the experimental setup} The setup consists of a sunlight concentration module and an entangled-photon source. The sunlight concentration module collects direct sunlight via a $1~\mathrm{m} \times 1.4~\mathrm{m}$ Fresnel lens and concentrates the spectrally filtered light into a multimode fiber (MMF). In the entangled-photon source, the sunlight is collimated and filtered to 405 nm $\pm$ 0.75 nm (inset) to pump a ppKTP crystal inside a PSI. We have carefully balanced the PSI to compensate for most spatial and temporal distinguishabilities between the SPDC processes in two directions. The resulting SPDC field is split and detected by MMF-coupled APDs to characterize two-photon entanglement and test for violation of local realism. }
\end{figure}

\section{Experimental setup}\label{sec2}

Figure~\ref{fig:1} depicts the experimental setup, which consists of a sunlight-concentration module and an entangled-photon source. The sunlight concentration module is designed to optimally isolate sunlight in the desired spectral bandwidth and couple them into the MMF. To this end, we first pre-focus the sunlight using a rectangular (1~m $\times$ 1.4~m) Fresnel lens covered with commercial-grade color films that partially block the warm-colored spectral components. This beam then propagates through a short-pass filter and a dichroic mirror to further reduce the optical power outside of the ultraviolet spectral band. The focused and filtered sunlight is then coupled into a MMF with a core diameter of 50~$\mu$m and numerical aperture (NA) of 0.22 (MMF-50-0.22) using a custom glass conic concentrator. Additional details relevant to the sunlight concentration module are included in the Supplemental Document.

To efficiently couple the sunlight into the nonlinear medium responsible for inducing SPDC, we first collimate the sunlight transmitted through the MMF into a beam using a microscope objective with $20\times$ magnification and 0.4~NA. This beam, which will later serve as the pump of the SPDC processes, is spectrally filtered using a short-pass filter with a cutoff wavelength of 550 nm and a band-pass filter with a center wavelength of 405 nm and a bandwidth of 1.5 nm (BPF-405). Using a combination of a polarizing beam splitter (PBS) and a half-wave plate (HWP), we make the pump beam highly polarized with equal amplitudes in its horizontal and vertical components, which ensures the generation of high polarization entanglement~\cite{Kulkarni2016PRA}. To account for the overall phase delays that the orthogonal polarization components may later experience, we add a quarter-wave plate (QWP) that can be tilted to pre-compensate for the expected phase differences. The polarized sunlight pump beam is then focused into the nonlinear medium using a lens with a focal length of $f_1 = 200$~mm. The filtered sunlight reflected by the PBS is sent to a power meter to monitor the pump power in real time or to a spectrometer to measure its spectrum. Since sunlight is nearly unpolarized, and considering the depolarizing effect of propagation in MMFs, we expect the pump power to be equally split at the PBS so that the power meter readings reflect the actual pump power incident on the nonlinear medium. Similarly, we expect the PBS to have negligible influence on the spectral shape of light around 405 nm, so that the spectrum measured at the spectrometer reflects the spectrum of the actual pump beam that induces the SPDC process. We depict the spectrum of the filtered sunlight pump beam in the inset of Fig.~\ref{fig:1}. At the center of the nonlinear medium, the sunlight pump beam has a diameter of $\sim1$~mm with an estimated divergence half-angle of 11~mrad. 

The nonlinear medium is a 10-mm-long periodically poled potassium titanyl phosphate (ppKTP) crystal quasi-phase-matched for Type-II SPDC from 405~nm to~810 nm. By placing this crystal at the center of the PSI, we have made the SPDC processes induced in clockwise and counterclockwise directions indistinguishable by the detectors. As a result, the setup produces a polarization-entangled two-photon state $|\Psi\rangle \sim|HV\rangle+e^{i\phi}|VH\rangle$, where $\phi$ is jointly determined by the initial polarization of the pump beam and the subsequent phase delays acquired within the PSI \cite{Kim2006PRA}. In the experiment, we have set the tilting angle of QWP to target $\phi = \pi$ so that the resulting two-photon state is written as
\begin{equation}\label{eqn1}
    |\Psi\rangle = \frac{1}{\sqrt{2}}(|HV\rangle-|VH\rangle)
\end{equation}
The SPDC fields emitted into two distinct paths are each collimated by a lens with a focal length of $f_2 = 250$~mm. We denote the down-converted photons emitted into the upper path as \textit{signal} while those emitted into the lower path as \textit{idler}. Their joint polarization state is analyzed using a set of QWP, HWP, and PBS in each path. Note that we have chosen all waveplates in the setup to be zero-order (except for the dHWP inside the PSI) and have carefully aligned their optical axes before taking measurements. We use microscope objectives (OBJs) to couple down-converted photons into MMFs with core diameters of 200 $\mu$m and NAs of 0.39 (MMF-200-0.39) and detect them using avalanche photodiodes (APDs, PerkinElmer SPCM-CD 3017/Excelitas SPCM-QRH-14-FC). 

We register the photon arrival time using a time-to-digital converter (TDC, qutool's quTAU) with a time resolution of $\tau= 81$~ps and extract the time correlation histograms between signal and idler photons. To characterize the resulting two-photon polarization state, we perform quantum state tomography~\cite{James2001PRA} by recording coincidence rates at 16 different polarization projection bases and reconstruct the two-photon density matrix $\rho$, from which we can quantify the entanglement by calculating the concurrence $C(\rho)$~\cite{Wootters1998PRL} and purity $P(\rho) = \text{Tr}\{\rho^2\}$. We set an acquisition time of 2~min and a coincidence time window of 1~ns for each projective measurement. Due to solar irradiance variation throughout the daytime under realistic weather conditions, we observe differences in the sunlight pump power between measurements. Since the sunlight pump power measured at the power meter is well below 1~mW, we expect the SPDC to work in the low-gain regime where the pair generation rate scales linearly with the pump power. In subsequent data processing, we normalize the experimentally measured coincidence counts against the average pump power during data acquisition. The accidental coincidence rates are estimated using the average coincidence rates of photons arriving more than 5~ns apart. 

\section{Results and discussions}\label{sec3}

\begin{figure}
\includegraphics[width=\textwidth,height=\textheight,keepaspectratio]{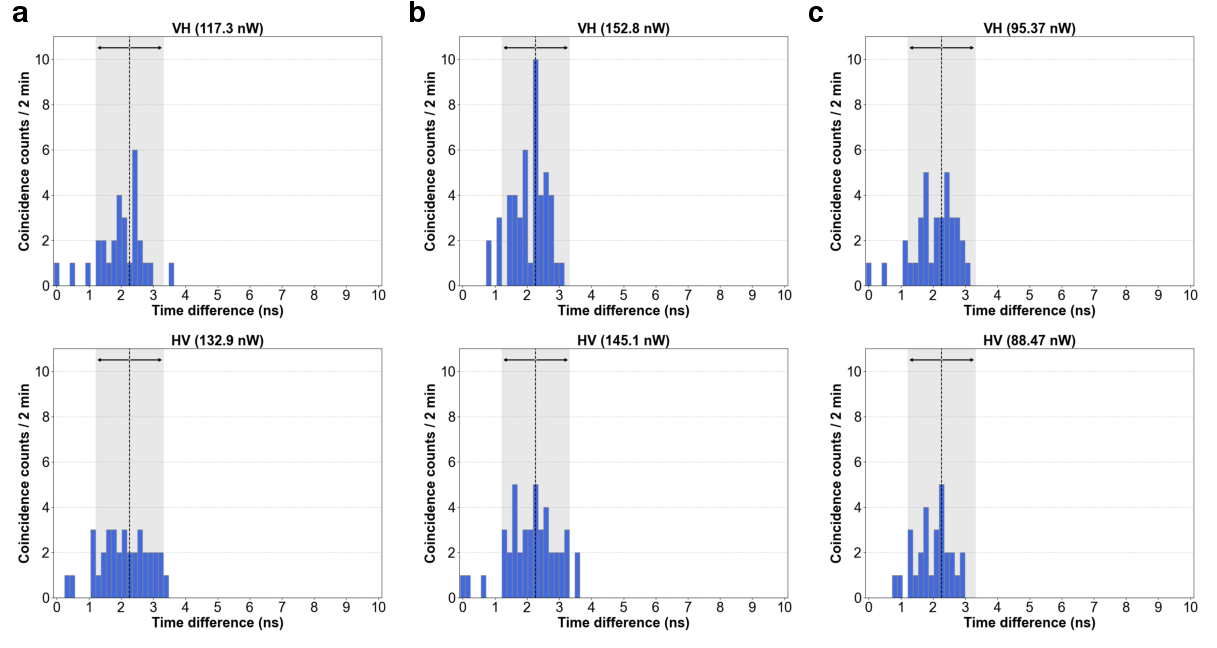}
\caption{\label{fig:2} \textbf{Time-correlation histogram of sunlight-pumped SPDC measured in joint polarization bases VH and HV, respectively} In the titles of subplots, we note the average pump power measured over the time of data acquisition. (a-c) display results measured on three separate days. The bar plots display the experimentally measured photon coincidences, while the grey shade indicates the coincidence time window. The photon pairs show clear temporal correlation with negligible accidental coincidences.}
\end{figure}

Figure~\ref{fig:2} displays the experimentally measured time correlation histogram between the SPDC photons measured on three separate days. We have set the waveplates in the detection paths to measure in the joint polarization bases VH and HV, respectively. We acquire data for 2 min per measurement and record the average pump power during the data acquisition. The time bin size is set to be twice the TDC's temporal resolution, $2\tau=162~$ps. We observe photon coincidences centered around $\Delta t\sim2.25$~ns. This time difference is caused by electronic delays between different channels of the TDC unit. At a 1~ns coincidence time window, we extract an average coincidence rate of 10 counts per minute per 100~nW of pump power, which is equivalent to a coincidence rate of $\sim 1600~\mathrm{s}^{-1}~(\mathrm{mW~of~pump~power})^{-1}$. This coincidence rate is consistent with the results obtained in an earlier work, in which a blue-light LED filtered to the same spatiotemporal bandwidth is employed as the pump source for SPDC and produces 100 counts per minute per 1~$\mu$W of pump power~\cite{li2025pra}. We also note that, when this result is normalized against the effective bandwidth of the nonlinear interaction, the pair generation rate from incoherent-light-pumped SPDC is comparable to that of laser-pumped SPDC reported in~\cite{li2025pra}. We also observe near-zero coincidence count rates at time differences far from the center of correlation, that is, $\Delta t > 5$~ns. Therefore, the experimentally measured coincidence counts contain negligible accidental coincidence counts. These results indicate a strong temporal correlation between the signal and idler photons, which is a feature of photon pair production from SPDC. Interestingly, an earlier study has observed spatial correlations between photon pairs produced from sunlight-pumped SPDC without studying entanglement~\cite{xing2025arxiv}. To observe photon entanglement beyond spatial and temporal correlations, it is necessary to reconstruct the density matrix of the quantum state and test for violations of Bell-type inequalities.

\begin{figure}
\includegraphics[width=0.98\textwidth,height=0.98\textheight,keepaspectratio]{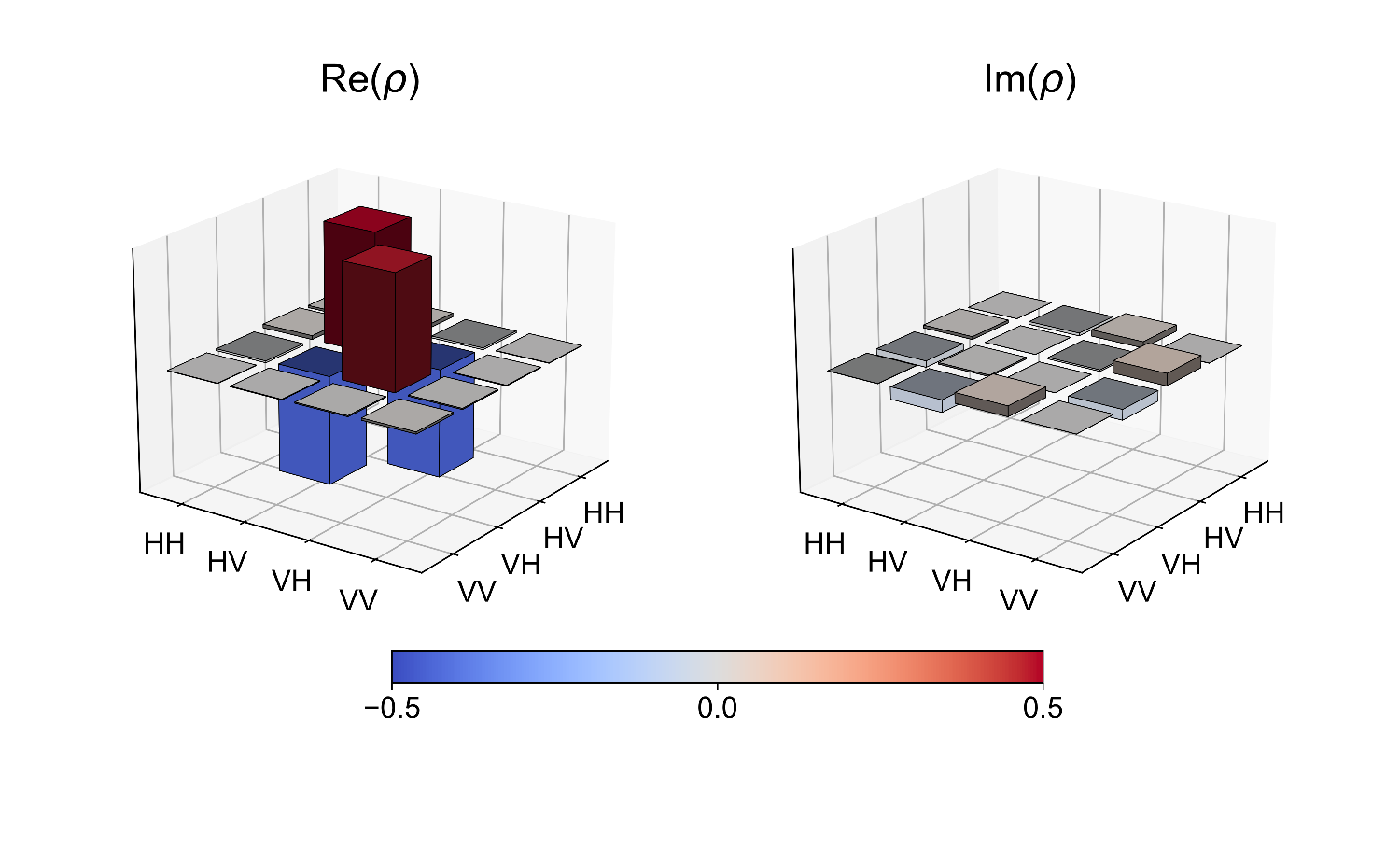}
\caption{\label{fig:3} \textbf{Real and imaginary parts of the density matrix of the two-photon state produced from sunlight-pumped SPDC.} These results are calculated using quantum state tomography based on polarization correlation measurements in 16 different projection bases, which include those that have been displayed in Fig.~\ref{fig:2}. These results confirm polarization entanglement with a concurrence of $C =0.905\pm0.053$, a purity of $P =0.919\pm0.045$, and a fidelity of $F = 0.939\pm0.027$ to the target Bell state (Eqn.~\ref{eqn1}).}
\end{figure}

Figure~\ref{fig:3} depicts the real and imaginary parts of the two-photon density matrix for sunlight-pumped SPDC. We obtain the depicted density matrix by averaging over the ones reconstructed from QST measurements conducted on three separate days. The resulting two-photon state has a concurrence of $C =0.905\pm0.053$.  Recall that the concurrence $C(\rho)$ is defined as an entanglement monotone that ranges from 0 to 1, where $C=0$ represents complete absence of entanglement and $C=1$ stands for maximal entanglement \cite{Wootters1998PRL}. Therefore, we have, for the first time, observed the generation of polarization-entangled photons from SPDC pumped by sunlight. To fully characterize the resulting state, we calculate the purity $P = \mathrm{tr\{\rho^2\}}$, where $\mathrm{tr\{\cdots\}}$ denotes the trace of a matrix, and the fidelity $F$, which quantifies the overlap between the generated state and the maximally entangled target state (Eqn.~\ref{eqn1}). As a result, we obtain $P =0.919\pm0.045$ and $F = 0.939\pm0.027$. These already high values for the concurrence, purity, and fidelity indicate that the state is highly entangled, mostly pure (low noise), and close to the desired Bell state and are likely not limited in any fundamental way by the sunlight-pumped SPDC, but rather a consequence of optimizable technical parameters. Specifically, we believe that the spatial incoherence of the sunlight pump does not fundamentally degrade the entanglement quality in the polarization degree of freedom, and further improvements in entanglement quality can be achieved with an optimized setup design, such as using optical components with low wavefront distortions~\cite{Li2025arxiv1, li2025pra}.

\begin{figure}
\includegraphics[width=\textwidth,keepaspectratio]{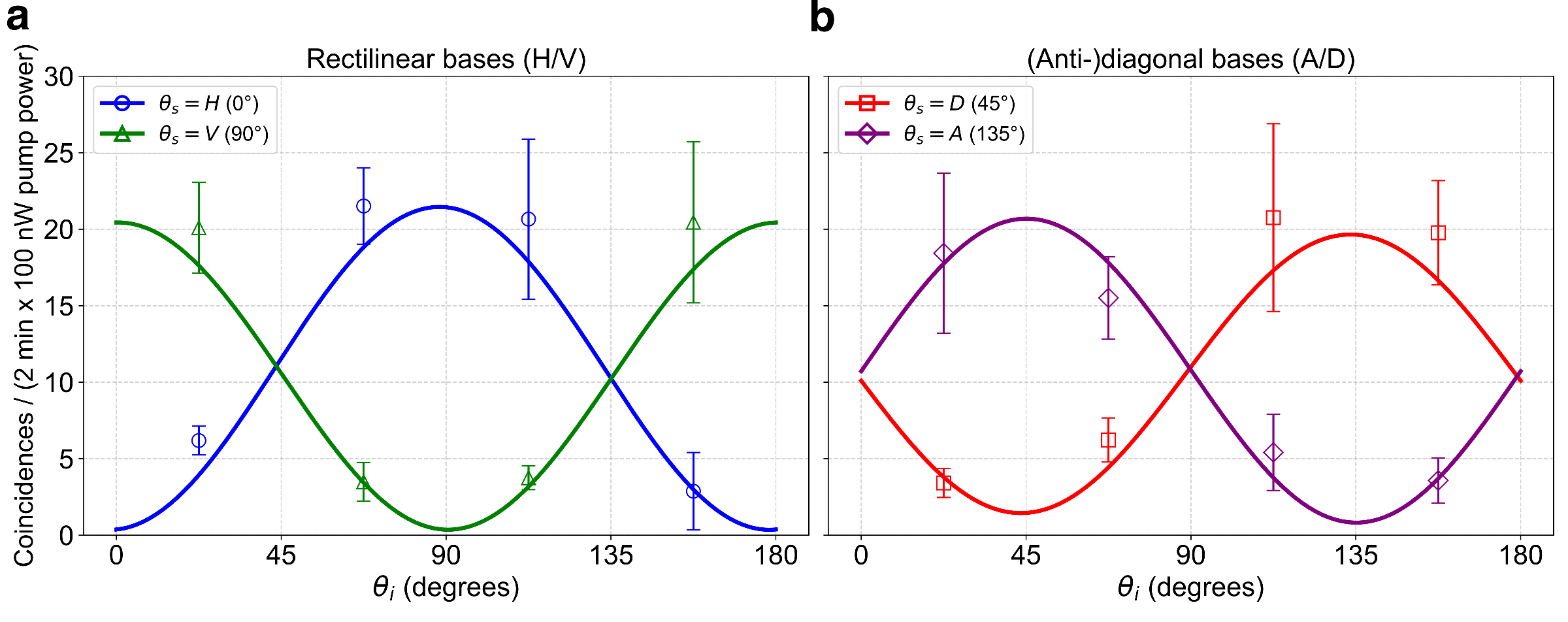}
\caption{\label{fig:4} \textbf{Bell parameter measurement using the CHSH formalism.} Idler photons are projected into different linear polarizations while signal photons are projected into the \textbf{(a)} rectilinear bases and \textbf{(b)} (anti-)diagonal bases. We denote $\theta_{s(i)}$ as the angle of linear polarization with respect to the horizontal polarization, so that $\theta_{s} = 0^\circ, 45^\circ, 90^\circ, 135^\circ$ represents projecting signal photons in the horizontal (H), diagonal (D), vertical (V), and anti-diagonal (A) bases, respectively. These results lead to a violation of local realism with a Bell-CHSH parameter of $S =  2.5408 \pm 0.2171 > 2$.}
\end{figure}

While the tomography results quantify the quality of entanglement, verifying the non-classical nature of these correlations requires a test against local realism. Violation of local realism arguably distinguishes quantum systems from their classical counterparts \cite{Einstein1935PhysRev, Bell1964, Clauser1969PRL} and could thus benchmark the performance of quantum systems for practical applications. For instance, observing violations of local realism certifies the security of quantum communication channels \cite{Bennett1984, Ekert1991PRL, Mayers1998} and serves as a valuable resource for quantum-enhanced imaging and metrology \cite{Moreau2019sciadv, Niezgoda2021PRL, Yadin2021ncomm}. To demonstrate this capability with our sunlight-pumped entangled-photon source, we perform a Bell test using the CHSH formalism.

In Figure~\ref{fig:4}, we depict the experimentally measured photon coincidences for a Bell test in the CHSH formalism. To account for pump power fluctuations between different measurements, we have normalized the coincidence counts to a reference pump power of 100~nW before further data processing. The markers and error bars indicate the experimentally measured average coincidences and their standard deviations over measurements conducted on three separate days. We have used these data to calculate the Bell-CHSH parameter. The solid lines represent the expected polarization correlation curves, which are numerically calculated by projecting the resulting two-photon density matrix shown in Fig.~\ref{fig:3} onto different joint polarization bases. The results of Bell test measurements agree well with those reconstructed from state tomography, indicating high consistency of the setup performance. We choose $\theta_{s} = 0^\circ, \theta_{s}' = 45^\circ, \theta_{i} = 22.5^\circ, \theta_{i}' = 67.5^\circ$ and calculate the CHSH parameter $S$ from the coincidence rates using:
\begin{equation}\label{eqn2}
    S = |E(\theta_s, \theta_i)-E(\theta_s, \theta_i')+E(\theta_s', \theta_i)+E(\theta_s', \theta_i')|,
\end{equation}
and $E(\theta_s, \theta_i)$ is defined as
\begin{equation}\label{eqn3}
    E(\theta_s, \theta_i)=
    \frac{N(\theta_s, \theta_i)+N(\theta_s^\perp, \theta_i^\perp)-N(\theta_s, \theta_i^\perp)-N(\theta_s^\perp, \theta_i)}{N(\theta_s, \theta_i)+N(\theta_s^\perp, \theta_i^\perp)+N(\theta_s, \theta_i^\perp)+N(\theta_s^\perp, \theta_i)},
\end{equation}
\begin{equation}
     \theta^\perp =  \theta + 90^\circ,
\end{equation}
where $\theta_{s(i)}^\perp$ stands for an angle perpendicular to $\theta_{s(i)}$. As a result, sunlight-pumped SPDC photons exhibit $S =  2.5408 \pm 0.2171 > 2$, thereby violating the CHSH inequality by 2.49 standard deviations. 

\section{Conclusion and Outlook}\label{sec4}

In summary, we have experimentally demonstrated the generation of polarization entanglement from SPDC pumped by sunlight. We observe a strong time correlation between photons produced from sunlight-pumped SPDC. Polarization correlations between the SPDC photons violate the Bell-CHSH inequality with $S =  2.5408 \pm 0.2171$, which is larger than the local realistic threshold of 2, and thus confirms the non-classical nature of these correlated photons. Tomography analysis shows that the two-photon state produced from sunlight-pumped SPDC has a concurrence of $C =0.905\pm0.053$, a purity of $P =0.919\pm0.045$, and a fidelity of $F = 0.939\pm0.027$ to the target Bell state (Eq.~\ref{eqn1}), confirming, respectively, the generation of high polarization entanglement, with low decoherence and in the desired state from sunlight-pumped SPDC. These results mark the first demonstration of an entangled photon source operating without electrical energy input (except for temperature control on the nonlinear crystal to maintain phase-matching conditions), thus paving the way for energy-efficient and environmentally friendly quantum entangled sources. Furthermore, we emphasize that the quality of entanglement generated from sunlight-pumped SPDC, as quantified by its concurrence, purity, and fidelity to the target state, is on par with what has been reported in recent works employing laser-pumped SPDC~\cite{Lohrmann2020APL, Lee2021QST, Brambila2023OE, Park2025apr}, highlighting the application potential of sunlight-driven entangled-photon sources in photonic quantum technologies.

Having measured high entanglement for these sunlight-pumped SPDC photon pairs, as well as confirming that the correlations are fundamentally non-classical, we next consider what technical alterations would further improve these figures of merit. 

One possible improvement would be to mitigate wavefront distortions in the optical components comprising the PSI. For instance, an uneven surface of the dPBS could introduce path distinguishability within the PSI and reduce the overall entanglement~\cite{Kim2006PRA, li2025pra}. This wavefront distortion could be characterized and compensated for using adaptive optics methods. Alternatively, one can employ optical components custom-designed for low wavefront distortion. We believe these improvements can significantly enhance the quality of the resulting entanglement, yielding near-unity concurrence, purity, and fidelity to a maximally entangled state.

Another question is how to improve the photon pair production efficiency. We have shown here that sunlight-pumped SPDC produces photon pairs at a rate of $\sim 1600~\mathrm{s}^{-1}~(\mathrm{mW~of~pump~power})^{-1}$, which is the same efficiency observed with an LED of the same spectral profile \cite{li2025pra}, that is, 1.5 nm bandwidth centred around 405 nm. However, this is still somewhat less than the efficiency shown in the same setup for laser pumping ($\sim 7500~\mathrm{s}^{-1}~(\mathrm{mW~of~pump~power})^{-1}$). This discrepancy is likely due to the limited spectral phase-matching bandwidth of the nonlinear crystal (0.2-0.3 nm centred around 405 nm), which is much wider than the laser bandwidth, meaning that only a fraction of the power in the input sunlight participates in the SPDC and produces entangled photons. The spectral utilization efficiency can be improved by adopting type-0 phase-matching. In this experiment, we have chosen the MMF-50-0.22 to maximize the overlap between the transverse profile of the sunlight pump beam and that of the ppKTP crystal in the same plane. By adopting a nonlinear medium with larger transverse dimensions, the SPDC processes can potentially accommodate an increased amount of pump power distributed over a wider spatial spectrum.

Overall, we stress that entanglement quality and production efficiency are currently only limited by technical challenges and do not suggest a fundamental difference in the quality of an entangled photon source with an incoherent pump.

To further improve the overall brightness of entangled photon generation, one should refine the design of the solar concentration module for better sunlight collection efficiency. We identify several technical aspects to be improved in this avenue. Firstly, one can optimize the design of the conic concentrator so that its effective clear aperture and acceptance angle better match the focal parameters of the Fresnel lens at the desired spectral band. Wavelength-specific optical coatings could also be applied to the concentrator to minimize reflection and scattering losses. Secondly, one can adopt a motorized solar-tracking mount with finer angular resolution so that the overall coupling efficiency of sunlight into the MMF is less susceptible to motor inaccuracies and environmental disturbances. Thirdly, one can explore a filtering element engineered for high transmission at the pump wavelength to directly increase the usable optical flux. Although the commercial-grade color film employed in the current setup performs well at rejecting light at longer wavelengths, it also reflects a significant fraction of the light near 405 nm (see Supplemental Document), thereby reducing the available pump power. Finally, one can employ collection fibers with a larger core diameter. This would not only enhance the coupling efficiency of concentrated sunlight but, when paired with an appropriate lens imaging system, also deliver more overall pump power to a nonlinear crystal with larger transverse dimensions, thereby facilitating more effective utilization of the spatially diverse solar pump. 

Although several opportunities for technical optimization have been identified, the highly entangled photons demonstrated in this proof-of-principle experiment already make a compelling case for replacing power-hungry, high-maintenance lasers with energy-efficient sources such as sunlight or LEDs. This technology offers strategic advantages, particularly for deploying quantum technologies in resource-restricted environments such as the Arctic or satellites in space \cite{Yin2020Nature}. 

Beyond logistical benefits, entangled photons generated from incoherent sources possess unique performance advantages. For instance, sunlight-pumped SPDC could take advantage of the broad spectral band of solar irradiance to facilitate access to entangled photons over a wider spectral range. Combined with the inherent low coherence between spectral components, these states may have optimized robustness against cross-talk between adjacent wavelength channels and thus offer a larger information capacity \cite{Dong2024Nature}. Additionally, studies indicate that polarization-entangled photons pumped by spatially partially coherent light display stronger resilience to atmospheric turbulence, making them highly suitable for free-space quantum key distribution \cite{Bhattacharjee2020OptLett, Qiu2012APB, Phehlukwayo2020PRA}. Interestingly, recent findings show that four-wave mixing driven by amplified spontaneous emission can produce highly entangled states with higher generation rates than those driven by coherent emission \cite{Song2025}. These findings not only challenge the conventional presumptions on the relations between coherence and entanglement but also present a new opportunity for boosting the energy efficiency and performance of current quantum information technologies.

\begin{backmatter}
\bmsection{Author contributions}
C.L., J.U., H.F., and R.W.B. conceived the research idea. J.B. and M.K. designed and built the solar concentration module. C.L. built the entangled-photon source. C.L. and J.B. conducted outdoor experiments with M.K.'s help to measure sunlight-driven entangled photon generation. C.L., J.B., J.U., and H.F. analyzed the data. C.L. and J.B. wrote the first draft of the paper, and all authors subsequently contributed to the writing. H.F. and R.W.B. supervised the project.

\bmsection{Funding}
The portion of the work performed at the University of Ottawa was supported by the Canada Research Chairs program under Award 950-231657, the Natural Sciences and Engineering Research Council of Canada under Alliance Consortia Quantum Grant ALLRP 578468 - 22, Discovery Grant RGPIN/2017-06880, and the Canada First Research Excellence Fund Award 072623. The work performed at the Max Planck Institute for the Science of Light was supported by the Max Planck Society and a scholarship from Erlangen School in Advanced Optical Technologies. In addition, R.W.B. acknowledges support from the U.S. National Science Foundation Award No. 2138174 and the U.S. Department of Energy Award No. FWP 76295. 

\bmsection{Acknowledgement}
The authors acknowledge useful discussions with Prof. Wuhong Zhang and Prof. Maria Chekhova. The authors thank Prof. Jeff Lundeen, Dr. Manuel Francisco Ferrer García, Prof. Christoph Marquardt, Thomas Dirmeier, and Dr. Yen-Ju Chen for access to key experimental equipment and software.

\bmsection{Data availability}
All relevant data are available from the corresponding author upon reasonable request.

\bmsection{Disclosures}
The authors declare no conflict of interest.

\end{backmatter}
\bibliography{main}

\end{document}


\begin{abstract}
\end{abstract}
\maketitle

\section{Outdoor experimental setup}
Fig. S1 depicts the actual experimental setup, which is deployed in an enclosed outdoor space at the Max Planck Institute for the Science of Light in Erlangen, Germany. To minimize the influence of wind, the experimental area is surrounded by protective fencing. The full setup for generating entangled photons, together with the APDs, TDC, and the temperature control unit of the ppKTP crystal, is placed inside a tent. This outer enclosure provides a darkroom-like environment that facilitates onsite troubleshooting of optical alignment for a setup placed outside a traditional laboratory. Within this tent, the entangled-photon source setup is enclosed in a light-tight housing that prevents any residual external light from entering the system. The APDs are mounted inside this inner enclosure to minimize the detection of background illumination and accidental coincidences. The pre-filtered and concentrated sunlight is guided to the entangled-photon source through a five-meter MMF with a core diameter of 50~$\mu$m and a numerical aperture (NA) of 0.22 (MMF-50-0.22). One end of this fibre is butt-coupled to the sunlight concentration module, while the other end is routed through a small opening at the bottom of the tent, and connected to the OBJ (with 20$\times$ magnification and 0.4~NA) that couples the light into the free-space entanglement setup.

\begin{figure}[ht]
    \centering
    \includegraphics[width=\textwidth]{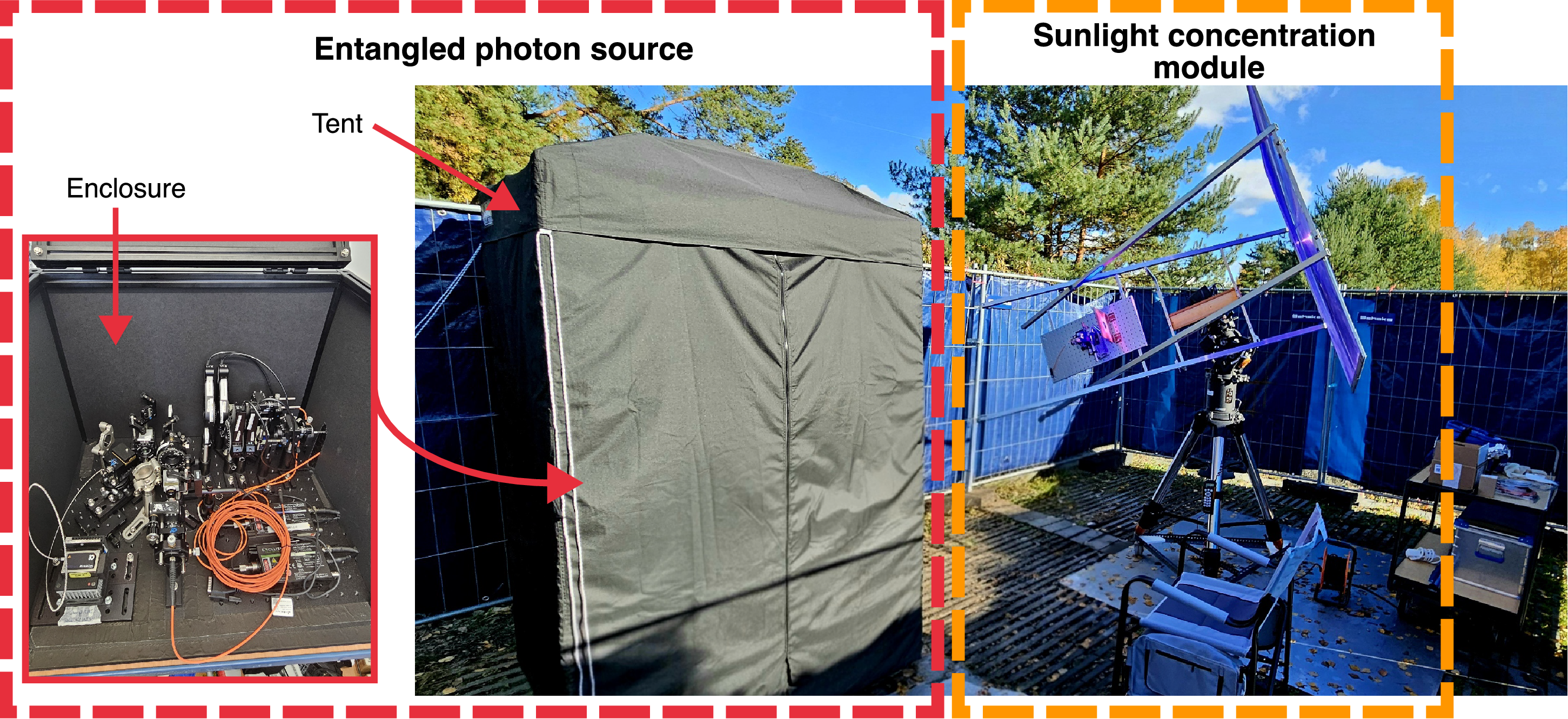}
    \caption{\textbf{Real-life photographs of experimental setups used for the sunlight-pumped generation of polarization-entangled photon pairs} The entire setup consists of two components: a sunlight concentration module (right) and an entangled-photon source (left) housed inside a protective and darkening tent, which also contains the associated electronics. Inset: Optical setup of the entangled-photon source, showing the optical arrangement and avalanche photodiodes (APDs) used for photon-pair generation. A hinged door (not shown) is also installed as part of the enclosure, which, when closed, covers the front and top sides of the optical setup.}
    \label{fig:S1}
\end{figure}

\section{Sunlight concentration system}

\begin{figure}[ht]
    \centering
    \includegraphics[width=0.9\textwidth, keepaspectratio]{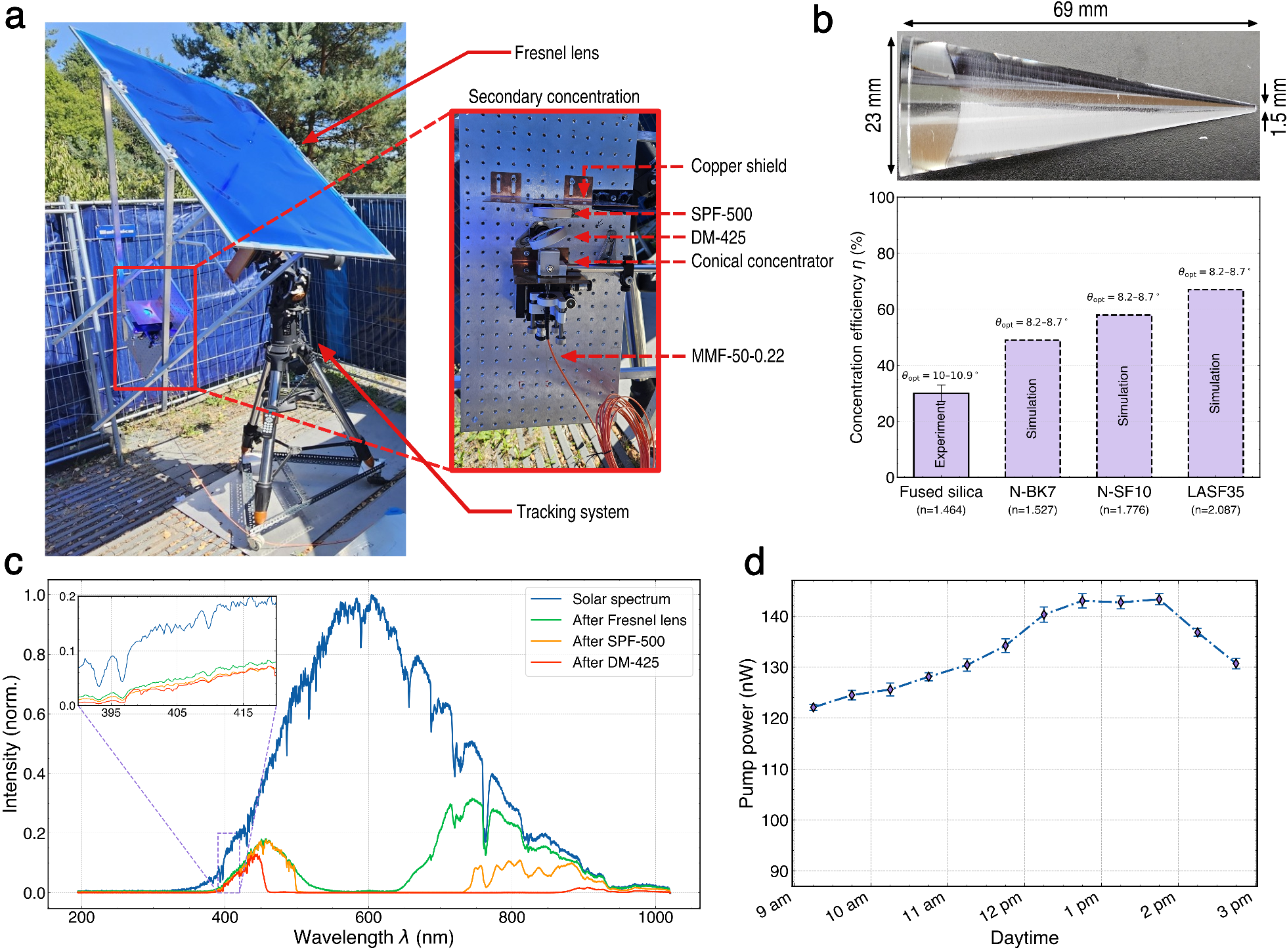}
    \caption{\textbf{Detailed schematics of the sunlight concentration module and its performance} (\textbf{a}) Real-life photograph of the outdoor solar concentration setup showing the Fresnel lens serving as the primary concentrator, mounted on a dual-axis tracking system, together with the secondary concentration module. Inset: secondary concentration module. (\textbf{b}) Geometry and dimensions together with a plot indicating the optimization parameters of the custom-designed conical concentrator used to efficiently couple concentrated sunlight into the multimode fibre.  (\textbf{c}) Experimentally measured spectra at successive stages of the solar concentration, illustrating the gradual spectral filtering of sunlight. They are normalised and scaled to the respective measured power, summed over the entire spectrum. Inset: spectrum around target pumping wavelength 405~nm (\textbf{d}). Diurnal variation of the pump power over the course of the day under clear-sky conditions, showing a maximum near solar noon. The data were recorded on 18 September 2025. SPF-500: short-pass filter with a cut-off wavelength of 500~nm; DM-425: dichroic mirror transmitting wavelengths shorter than 425~nm and reflecting longer wavelengths; MMF-50-0.22: multimode fiber with a core diameter of 50~$\mu$m and a numerical aperture of 0.22.}
    \label{fig:S2}
\end{figure}

Fig. S2 displays the design and performance of the sunlight concentration module used for solar pumping. Fig. S2(a) shows a photograph of the actual sunlight concentration setup. Incident sunlight is collected by a 1~m $\times$ 1.4~m PMMA Fresnel lens serving as the primary concentrator, focusing the light to a focal spot of 13~mm in diameter at a focal length of 90~cm. To reduce the contribution of long-wavelength spectral components, the Fresnel lens is covered with commercial-grade colour filters (Roscolux). The lens, clamped between two arms of a supporting frame, is mounted on a high-precision equatorial mount (Celestron CGE Pro). Following polar alignment to the celestial north pole, the Sun’s position is calculated using software integrated into the mount, which automatically slews the system to the solar position and provides continuous motor-driven tracking. The mount achieves a tracking accuracy of approximately 9~arcseconds, ensuring stable alignment of the Fresnel lens with the sun throughout the measurement period. Residual tracking errors and environmental perturbations nevertheless introduce slow fluctuations in the focal-spot position at the secondary optics, which limit the overall coupling efficiency into the subsequent concentrator and fiber. Employing a motorized solar-tracking mount with finer angular resolution would reduce sensitivity to such effects and improve long-term coupling stability, allowing increased acquisition times (currently on the order of ~2 min per measurement).

After being pre-filtered by the films on the Fresnel lens, the sunlight then propagates through additional secondary filters located at the focal plane of the Fresnel lens. This secondary concentration module is protected by a copper shielding structure with a circular aperture that is concentrically aligned with subsequent optical components. We have designed this copper shield to maximize sunlight throughput while effectively protecting the subsequent optical elements from being heated up by stray radiation. The beam first passes through a short-pass filter with a cut-off wavelength of 500~nm (SPF-500, Edmund optics, 84-719), and subsequently through a dichroic mirror transmitting wavelengths shorter than 425~nm and reflecting longer wavelengths (DM-425, Thorlabs DMLP425L) at a 45° angle of incidence. Fine adjustment of the dichroic mirror angle with respect to the optical axis is used to maximize transmission at 405~nm. Through this multi-stage filtering process, the solar spectrum is progressively narrowed to the ultraviolet–blue wavelength range, selectively retaining the target pump wavelength of 405~nm. 

Fig. S2(c) depicts the experimentally measured transmission spectra after each stage of filtering, with the solar spectrum serving as a reference. The inset shows an enlarged view of the transmission behaviour around 405~nm. These are normalized and scaled to the respective full-spectrum power. As shown, the initially broadband solar spectrum is progressively attenuated at each filtering stage, i.e. the Fresnel lens, the SPF-500 and the DM-425, resulting in a corresponding reduction of the transmitted power. 

In order to efficiently couple the light focused and filtered by the Fresnel lens and secondary filters into a multimode fiber with a core diameter of 50~$\mu$m and a numerical aperture of 0.22 (MMF-0.5-0.22), a custom-made conical concentrator is used \cite{JasvinderThesis}. Its shape and geometry are shown in Fig. S2(b), along with a plot highlighting the relevant geometric and optical optimization parameters. The solid cylindrically symmetric conical concentrator is fabricated from fused silica. It has an overall length of 69~mm, an entrance aperture diameter of 23~mm and an exit aperture diameter of 1.5~mm. Light entering the wide aperture is guided toward the narrow output by refractive guiding and total internal reflection at the silica–air interface, resulting from the tapered geometry. This design enables efficient concentration of incident solar radiation onto a small area while maintaining high optical transmission and thermal stability. Experimentally, a concentration efficiency $\eta$ of approximately 30 $\pm$ 3\% is obtained. This efficiency is defined as the ratio of the output power measured at the smaller end of the cone to the input power at the wider end. Importantly, the existing conical concentrator was originally designed to maximize collection efficiency across a broad visible spectral range, rather than being optimized specifically for narrowband collection at 405~nm. As a result, additional improvements in the usable 405~nm throughput are expected through the application of wavelength-specific anti-reflection coatings, refinement of the acceptance angle to better match the focused solar cone, and optimization of the entrance aperture relative to the Fresnel-lens focal spot \cite{SolarPumpedLaser2025,JasvinderThesis}.

Ray tracing simulations indicate that, with the appropriate choice of glass material and further geometric and optical optimizations, this concentrator can achieve concentration efficiencies of up to 67\% \cite{JasvinderThesis}. The plot in Fig. S2(c) compares the maximum concentration efficiencies obtained for four different glass materials at their respective optimal opening angles, denoted as $\theta_{\text{opt}}$, which correspond to the opening angles yielding the highest efficiency. We define the opening angle $\theta$ as the half-angle between the cone axis and the tapered sidewall of the concentrator, thereby directly describing the degree of tapering of the conical geometry. In contrast to fused silica ($n_{405} = 1.464$) used in the experiment, simulations for N-BK7 ($n_{405} = 1.527$), N-SF10 ($n_{405} = 1.776$), and LASF35 ($n_{405} = 2.087$) predict increasing efficiencies of 49~\%, 58~\%, and 67~\%, respectively, all within a common optimal opening-angle range of $8.2^\circ-8.7^\circ$. These results demonstrate a clear trend of increasing concentration efficiency with higher refractive index, accompanied by a convergence of the optimal opening angle toward smaller values for higher-index glasses. 

Based on the latitude of Erlangen (49.6$^\circ$N) and season of measurement, the maximum solar power collectable by the 1.4~$\mathrm{m}^{2}$ Fresnel lens is approximately 1.1–1.2~kW, of which we estimate that approximately 1.7~W of this power lies within the $405\pm0.75$~nm spectral band. Due to the combined effect of spectral filtering and the light-concentrating properties of the cone, the full-spectrum optical power available at the tip of the conic concentrator can reach up to 5~W under near-perfect conditions. After coupling into the MMF, the entangled-photon source receives an optical power of 100–200~nW within a 1.5-nm bandwidth around the center wavelength of 405~nm. As discussed in the main text, this overall sunlight utilization efficiency into the setup is constrained by the spatial and spectral bandwidth of the nonlinear conversion process and can be further improved by optimizing the crystal geometry and choosing a broadband phase-matching condition. Additionally, the coupled sunlight power is significantly influenced by the weather conditions at the time of measurement and by the time of day. Fig. S2(d) shows an example of the diurnal variation of the pump power measured on 18 September 2025. The pump power increases from the morning hours to a maximum around solar noon before decreasing again toward the late afternoon. On the measurement days corresponding to the data presented in this paper, pump powers of up to 195~nW were achieved. Although sufficient pump power is available throughout the day to obtain meaningful statistics for sunlight-driven entangled-photon generation, measurements performed near solar noon yield higher count rates and reduced statistical uncertainties.



\clearpage 

\bibliography{supp_bib.bib}